\documentclass[reprint,superscriptaddress,amsmath,amssymb,aps,prl,a4paper]{revtex4-1}

\usepackage{times,amssymb,graphicx,siunitx,multirow,xcolor}
\usepackage{dcolumn}
\usepackage{bm}
\usepackage{datetime} 

\begin{document}
\title{Superfluorescence, free-induction decay, and four-wave mixing:\\ propagation of free-electron laser pulses through a dense sample of helium ions}
\newcommand{\tauSF}{\ensuremath{\tau_{\textrm{SF}}}~}
\newcommand{\tauD}{\ensuremath{\tau_{\textrm{D}}}~}
\newcommand{\term}[3]{#1\hspace{1pt}$^2$#2$_{#3/2}$}
\newcommand{\textcolour}[1]{\textcolor{#1}}

\author{James R Harries}%
\affiliation{QST, SPring-8, Kouto 1-1-1, Sayo, Hyogo, 679-5148 Japan}%

\author{Hiroshi Iwayama}%
\affiliation{UVSOR, IMS, Nishigo-Naka 38, Myodaiji, Okazaki, Aichi 444-8585, Japan}%
\affiliation{SOKENDAI, Nishigo-Naka 38, Myodaiji, Okazaki, Aichi 444-8585, Japan}%

\author{Susumu Kuma}
\affiliation{Atomic, Molecular, and Optical Physics Laboratory, RIKEN, Saitama, 351-0198 Japan}%

\author{Masatomi Iizawa}
\author{Norihiro Suzuki}
\author{Yoshiro Azuma}
\affiliation{Sophia University, Tokyo 123-4567 Japan}%

\author{Ichiro Inoue}
\affiliation{RIKEN SPring-8 Centre, Kouto 1-1-1, Sayo, Hyogo, 679-5148 Japan}%

\author{Shigeki Owada}
\affiliation{Japan Synchrotron Radiation Research Institute, Kouto 1-1-1, Sayo, Hyogo 679-5198, Japan}

\author{Tadashi Togashi}
\author{Kensuke Tono}
\affiliation{Japan Synchrotron Radiation Research Institute, Kouto 1-1-1, Sayo, Hyogo 679-5198, Japan}

\author{Makina Yabashi}
\affiliation{RIKEN SPring-8 Centre, Kouto 1-1-1, Sayo, Hyogo, 679-5148 Japan}%

\author{Eiji Shigemasa}
\affiliation{UVSOR, IMS, Nishigo-Naka 38, Myodaiji, Okazaki, Aichi 444-8585, Japan}%
\affiliation{SOKENDAI, Nishigo-Naka 38, Myodaiji, Okazaki, Aichi 444-8585, Japan}%

\date{\today}
\begin{abstract}
We report an experimental and numerical study of the propagation of free-electron laser pulses (wavelength \SI{24.3}{\nano\metre}) through helium gas. Ionisation and excitation populates the He$^{+}$~4$p$ state. Strong, directional emission was observed at wavelengths of \SIlist{469;164;30.4;25.6}{\nano\metre}. We interpret the emissions at \SIlist{469;164}{\nano\metre} as 4$p$-3$s$-2$p$ cascade superfluorescence, that at \SI{30.4}{\nano\metre} as yoked superfluorescence on the 2$p$-1$s$ transition, and that at \SI{25.6}{\nano\metre} as free-induction decay of the 3$p$ state.
\end{abstract}

\maketitle

\begin{figure}
	\includegraphics{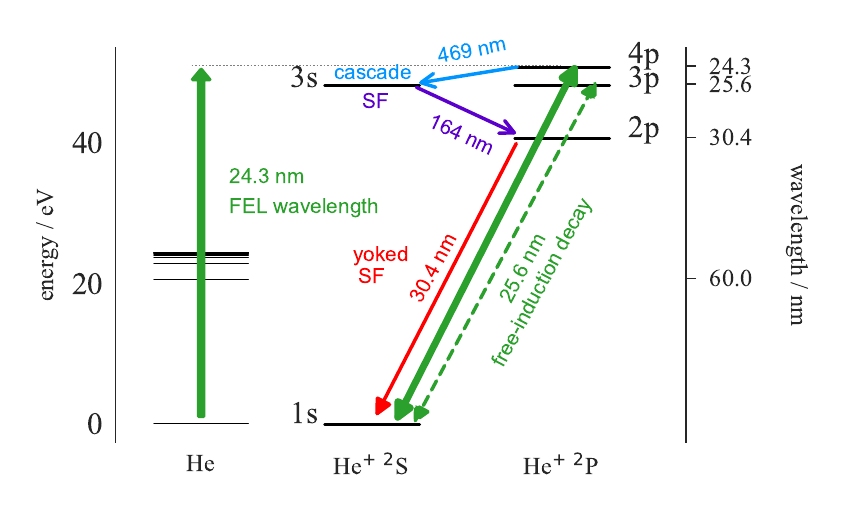}%
	\caption{\label{fig:level-diagram}Partial level scheme. The right-hand axis shows wavelength for transition to the ground state, the left-hand axis energy above the ground state for each system. The FEL pulses at a central wavelength of \SI{24.3}{\nano\metre} (green) ionise neutral He and resonantly couple the He$^+$ ground and 4$p$ states.} 
\end{figure}

Free-electron lasers  (FELs) operating at extreme ultraviolet wavelengths and shorter are currently allowing us to explore new possibilities in the interaction of coherent radiation with matter~\cite{rohringer_atomic_2012,weninger_stimulated_2013,weninger_stimulated_2013-1,kimberg_stimulated_2016,yoneda_atomic_2015}. The study of the propagation of intense, short-wavelength, coherent pulses through dense atomic media is an essential stepping stone to new analytical techniques, and the extension of existing technologies commonplace at visible wavelengths to wavelength regimes where, for example, element- and site-specificity can drastically expand their fields of applicability~\cite{adams_X-ray_2013}. Effects of current interest include superfluorescence (SF) and superradiance~\cite{nagasono_observation_2011,chumakov_superradiance_2018}, amplified spontaneous emission~\cite{rohringer_atomic_2012,yoneda_atomic_2015}, stimulated Raman scattering~\cite{weninger_stimulated_2013-1,kimberg_stimulated_2016}, and free-induction decay~\cite{bengtsson_spacetime_2017}. Here we study the propagation of partially-coherent FEL pulses at a wavelength of \SI{24.3}{\nano\metre} through a dense sample of helium ions. Experimentally we observe intense, highly-directional emission at several different wavelengths. Numerical simulations which follow the propagation of the full electric field and the evolution of the atomic medium over a timescale of tens of picoseconds are compared with the results. We interpret our findings (see figure~\ref{fig:level-diagram}) as cascade superfluorescence (\SIlist{469;164}{\nano\metre}), yoked superfluorescence (\SI{30.4}{\nano\metre}), and free-induction decay (\SIlist{25.6;24.3}{\nano\metre}). To our knowledge superfluorescence at such short wavelengths has not been reported previously, although recently few-photon superradiance has been observed at X-ray wavelengths~\cite{rohlsberger_collective_2010,chumakov_superradiance_2018}.

			\begin{table}\caption{\label{tab:manylevels}Wavelengths, characteristic superfluorescence times, and threshold column densities for selected transitions in He$^{+}$~\cite{kramida_nist_2015}. The threshold excited atom column density is estimated by equating $\tau_{\mathrm{SF}}$ and T$_{\textrm{diff}}$~(see text) for a diameter of \SI{12.6}{\micro\metre}. Transitions from \term{4p}{P}{1} are omitted for brevity.}
				\begin{ruledtabular}
					\begin{tabular}
						{ccdddd}
						upper & lower & \multicolumn{1}{c}{$\lambda$$\div$\si{\nano\metre}} &%
						\multicolumn{1}{c}{$\sigma\tau_{SF}$$\div$(\si{\pico\second\per\square\nano\metre})} &%
						\multicolumn{1}{c}{$\sigma_{\textrm{th}}$$\div$\si{\per\square\nano\metre}} \\ \hline

						\multirow{4}{*}{\term{4p}{P}{3}} & \term{2s}{S}{1} & 121.5  & 3.7  & 0.8 \\
						~                                & \term{3s}{S}{1} & 468.7  & 0.78 & 0.69 \\
						~                                & \term{3d}{D}{3} & 468.7  & 69   & 61 \\
						~                                & \term{3d}{D}{5} & 468.7  & 7.6  & 6.7 \\[6pt]
						
						\multirow{2}{*}{\term{3s}{S}{1}} & \term{2p}{P}{3} & 164.1  & 9.2  & 2.9 \\
						                                 & \term{2p}{P}{1} & 164.0  & 4.6  & 1.4 \\[6pt]
					    
						\term{3d}{D}{5} & \term{2p}{P}{3} & 164.0 & 0.30 & 0.09 \\[3pt]
						\multirow{2}{*}{\term{3d}{D}{3}} & \term{2p}{P}{3} & 164.0  & 1.8  & 0.56 \\
							                            & \term{2p}{P}{1} & 164.0  & 0.36  & 0.11 \\[6pt]
							                            
						\term{2p}{P}{3} & \term{1s}{S}{1} & 30.4 & 0.91 & 0.05 \\
						
						
					\end{tabular}\end{ruledtabular}
				\end{table}

\begin{figure}
	\includegraphics{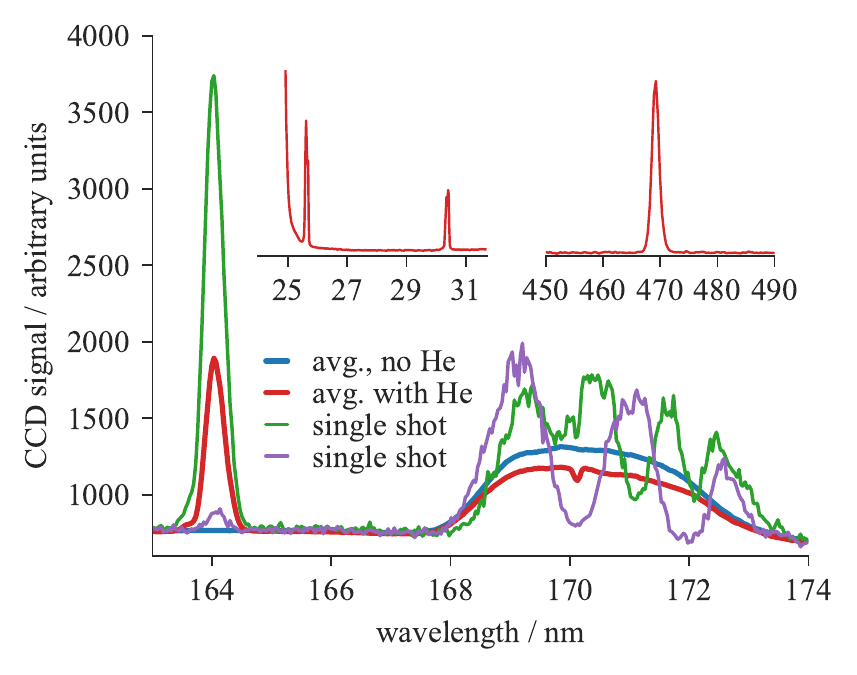}%
	\caption{\label{fig:spectrum164}Average and single-shot spectra recorded using the grazing-incidence spectrometer and CCD camera (main, left inset), and visible-wavelengths spectrometer (right inset). The incident FEL radiation (central wavelength \SI{24.3}{\nano\metre}) is seen in 7th grating order (\SI{170}{\nano\metre}), and the single-shot spectra reveal the multi-mode nature of the SASE pulses. }
\end{figure}

The experiments were carried out at SACLA BL1~\cite{owada_soft_2018,owada_single-shot_2018}. Pulses of central wavelength \SI{24.3}{\nano\metre}, energy around \SI{20}{\micro\joule}, and duration less than \SI{100}{\femto\second} were focussed to a spot size of around \SI{10}{\micro\metre} at the centre of a gas cell, into which helium gas was expanded using a pulsed nozzle (Parker). Two electrically-grounded apertures, of diameter \SI{1}{\milli\metre}, thickness \SI{2}{\milli\metre} and separation \SI{1}{\milli\metre} defined the interaction region. The ground-state number density was varied by varying the static pressure behind the pulsed nozzle. The gas cell was positioned at the centre of a cylindrical inner vacuum chamber (length \SI{230}{\milli\metre}, radius \SI{90}{\milli\metre}), pumped by a turbo-molecular pump. Two 10-mm-diameter apertures in this cylinder allowed the passage of the FEL beam and emitted radiation, and a further 15-mm-diameter aperture the passage of a gasline. The outer vacuum chamber was pumped by a second turbo-molecular pump, and was equipped with an ionisation gauge. Transmitted and emitted radiation was detected using either a flat-field grazing-incidence spectrometer (Shinkukogaku Co., Ltd. Japan) equipped with a CCD camera (Andor DV420), a visible wavelength spectrometer (Ocean Optics USB2000), or fast photodiodes (Hamamatsu G4176-03 with Picosecond pulse labs 550B-104 bias tee, Alphalas UPD-50-UP). The optical pathlengths from the centre of the gas cell to the detectors were around \SI{850}{\milli\metre} (CCD camera) and \SI{300}{\milli\metre} (photodiodes).

The main plot of figure~\ref{fig:spectrum164} shows example spectra recorded using the grazing-incidence spectrometer. The blue trace shows the average of 1000 single-shot spectra recorded with no He, and reveals (in 7th order of the grating, at \SI{170.1}{\nano\metre}) the average spectrum of the SASE pulses. With He present, (red trace), wavelength-independent absorption is seen due to ionisation. Using a value of \SI{1.9e-22}{\per\metre\squared} for the ionisation cross-section (interpolated from the results of Bizau and Wuilleumier~\cite{bizau_redetermination_1995}), an absorption length of \SI{5}{\milli\metre}, as defined by the outer faces of the two apertures of the gas cell, and assuming the ratio of emitted to incident radiation is given by the ratio of the areas under the broad spectral peaks (0.816), we can estimate a He density (and an upper limit He$^+$ density) of around \SI{2e23}{\per\metre\cubed}. Also evident in the red trace of figure~\ref{fig:spectrum164} is a narrow absorption dip at \SI{170.1}{\nano\metre} (\SI{24.3}{\nano\metre} in 7th order). This is due to resonant excitation of helium ions to the 4$p$ state. Strong emission is seen at a wavelength of \SI{164}{\nano\metre}, which corresponds to the n=3 to n=2 transitions in the ion (see figure~\ref{fig:level-diagram}). The example single-shot spectra reveal the multi-mode nature characteristic to the SASE FEL. From the (small) number of modes observed, it is possible to estimate that the pulse length is of the order of \SIrange{30}{100}{\femto\second}. A correlation between FEL intensity at \SI{24.3}{\nano\metre} and emission at \SI{164}{\nano\metre} is suggested. The inset on the right shows a spectrum recorded using the visible wavelengths spectrometer, revealing emission at \SI{469}{\nano\metre}. The inset on the left shows an average spectrum recorded using the grazing-incidence monochromator at shorter wavelengths, revealing strong emission at \SIlist{25.6;30.4}{\nano\metre}, and the tail of the broad FEL average spectrum.

The observation of emission at \SIlist{469;164}{\nano\metre} suggests cascade fluorescence on the route 4$p$-3$s$-2$p$ or 4$p$-3$d$-2$p$. While the branching ratio for the 4$p$-2$s$ transition for spontaneous fluorescence is around 0.11~\cite{kramida_nist_2015}, compared to around 0.04 for 4$p$-3$s$ and 0.004 for 4$p$-3$d$, emission at a wavelength of \SI{121.5}{\nano\metre} (4$p$-2$s$) was not observed, consistent with the interpretation that the emissions at \SIlist{469;164}{\nano\metre} are not due to spontaneous emision.

The characteristic superfluorescence time $\tau_{\mathrm{SF}}$ for a cylindrical sample (length $L \gg$ diameter $d$) can be estimated by $\tau_{\mathrm{SF}}=8\pi/(3\lambda_{ki}^2A_{ki}N_aL$)~\cite{gross_superradiance:_1982,benedict_super-radiance:_1996,allen_optical_1987}, where $\lambda$ is the wavelength of emission, $A$ the Einstein A coefficient, and $N_a$ the excited atom number density ($N_aL$ is then the excited atom column density $\sigma$). Both the width of the emitted pulse, and the characteristic delay following excitation are proportional to this value. Superfluorescence can develop on a particular transition only if its rate is greater than that of any other loss rate. For the experiments here, the most important loss rates are expected to be diffraction losses, and dephasing due to collisions with electrons. By equating $\tau_{\mathrm{SF}}$ to the diffraction loss rate 1/T$_{\mathrm{diff}}$ (estimated as $\lambda c/d^2$~\cite{ohae_production_2014,gross_superradiance:_1982}) it is possible to estimate a threshold excited atom column density $\sigma_{\textrm{th}}$ below which the development of superfluorescence is suppressed (see reference~\cite{harries_single-atom_2016-1} for details.) We estimate dephasing times due to electron collisions of a few tens of picoseconds~\cite{xia_observing_2012,hudis_x-ray-stimulated_1994}. In table~\ref{tab:manylevels}, we give the quantities $\sigma\tau_{\mathrm{SF}}$ and $\sigma_{\textrm{th}}$ for various transitions in He$^+$, assuming a diameter of \SI{12.6}{\micro\metre}, the estimated average diameter of the focussed FEL beam. For a given upper state, superfluorescence will preferentially develop on the transition with the fastest decay rate (and shortest delay), provided the threshold density is reached. For comparison, we estimate a column density of around \SI{200}{\per\nano\metre\squared} for ground state helium ions from the spectra of figure~\ref{fig:spectrum164} (see also the discussion in reference~\cite{harries_single-atom_2016-1}). From table~\ref{tab:manylevels} we can see that from the 4$p$ excited state the most likely superfluorescence transition is to the 3$s$ state, with transitions to the 3$d$ and 2$s$ states being slower and having higher thresholds. With sufficient population transfer to the 3s state, superfluorescence can be expected to proceed to the 2p state. The rate of the 3$s$-2$p$ transition is slower than that of the 4$p$-3$s$ transition, and its threshold is higher. Assuming no initial coherence in the excitation of the 4p state, the lack of population inversion prevents subsequent 2$p$-1$s$ superfluorescence. However, with coherent excitation the possibilty of \emph{yoked} superfluorescence~\cite{brownell_yoked_1995,ikeda_theory_1980,*ikeda_theory_1980-1} on this cascade scheme presents itself.

\begin{figure}
	\includegraphics{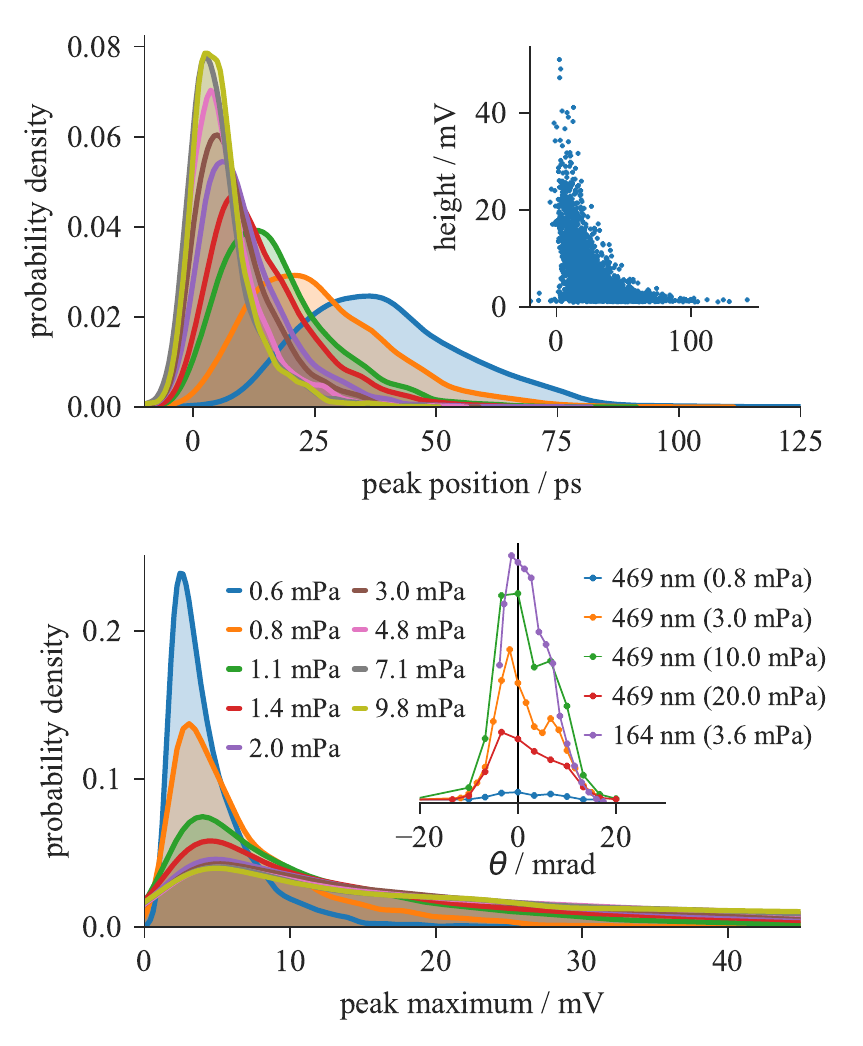}%
	\caption{\label{fig:469pulses}Properties of the emitted \SI{469}{\nano\metre} pulses. 1000 pulses were recorded at 9 different number densities. The upper plot shows the distributions of the peak position (with t=0 determined by fitting), and the lower plot the peak height, for all pulses with peak heights three times higher than the root-mean-square of the detector noise level. The inset to the upper plot shows the correlation between peak height and peak position for the traces recorded at a chamber pressure of \SI{0.8}{\milli\pascal}. The inset to the lower plot shows the angular dependence of the emissions (plot for \SI{164}{\nano\metre} arbitrarily scaled).
	}
\end{figure}

\begin{figure}
	\includegraphics{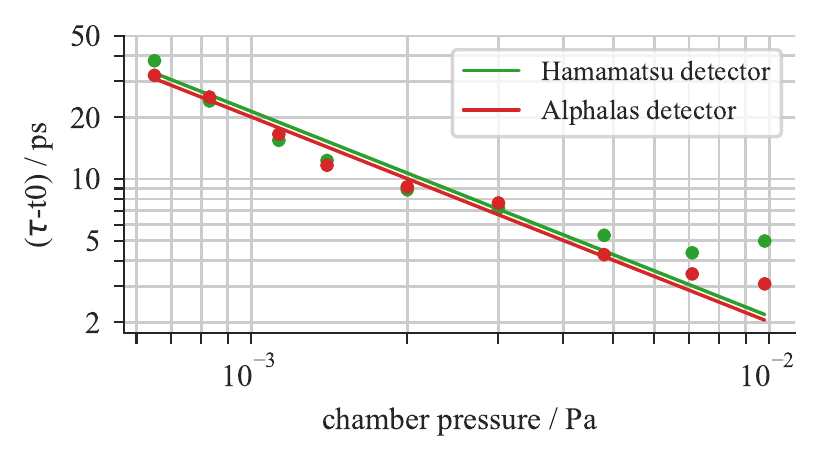}%
	\caption{\label{fig:delayfit}Median pulse delays (see figure~\ref{fig:469pulses}), and fits of the form $\tau=t_0+b/P$ (solid lines), where $P$ is the pressure in outer vacuum chamber, assumed to be proportional to the instantaneous ground state number density. The fitted $t_0$ parameters have been subtracted from the datapoints.
	}
\end{figure}

To confirm that the observed emission is indeed superfluorescence, and not spontaneous emission, we investigated the angular distribution of the fluorescence at \SIlist{469;164}{\nano\metre}, and the pulse delay of the emission at \SI{469}{\nano\metre} as a function of ground state atom density. For both wavelengths, angular diversions of less than \SI{10}{\milli\radian} were observed (see the lower inset to figure~\ref{fig:469pulses}). For comparison, the angular divergence of the incident FEL beam was less than \SI{1}{\milli\radian}. To investigate the pulse characteristics of the emitted radiation at \SI{469}{\nano\metre} individual pulses from the two photodiodes were recorded using an oscilloscope (Agilent DSO-X 91604A, bandwidth \SI{16}{\giga\hertz} ), using a signal correlated to the main FEL pulse as the trigger. Figure~\ref{fig:469pulses} shows the probability density distributions of peak height (lower plot) and peak position (defined as the point at which the signal reaches half-maximum) for 1000 pulses recorded at different nozzle backing pressures for the Hamamatsu detector. The labels correspond to the average pressures recorded in the outer vacuum chamber. The upper inset shows a scatterplot of height and position for one dataset. It is clear that higher number densities lead to larger peaks with shorter delays, a clear hallmark of superfluorescence. For these measurements, a beamsplitter was also used to direct a portion of the emission to the second photodiode (Alphalas UPD-50), which has poorer time resolution but showed similar results. Figure~\ref{fig:delayfit} shows the results of fits of the form $\tau=t_0+b/P$ ($\tau$ is peak position, P is chamber pressure) to the average peak positions for each dataset and detector. These results are consistent with the behaviour expected for superfluorescence. We suspect that the deviation at the highest pressures is due to a non-linearity between chamber pressure and the actual number density in the gas cell. The fitted values of the $b$ parameter were \SI{21}{\femto\second\per\pascal} for the Hamamatsu detector, and \SI{20}{\femto\second\per\pascal} for the Alphalas detector, and this difference can be interpreted as a measure of the systematic errors inherent to the setup and the measurements.

The (windowless) Alphalas detector has a small sensitivity at \SI{164}{\nano\metre}, and these measurements were also attempted for the \SI{164}{\nano\metre} emission, using a UV reflective filter (Acton M157) as a beamsplitter, and operating the detector in an atmosphere of nitrogen gas. No 164-nm-specific signal was observed with the FEL resonant with 4$p$ excitation, but weak pulses, with pulse widths indistinguishable from the detector response (less than \SI{200}{\pico\second}), were observed with the FEL resonant with 3$p$ excitation. This is consistent with superfluorescence on the 3$p$-2$s$ transition following 3$p$ excitation, however the signal was too weak to analyse signal delay times.

Extending our previous work~\cite{ohae_simultaneous_2016, harries_single-atom_2016-1}, and going beyond previous few-level studies~\cite{liu_dynamics_2008,sun_propagation_2010,krusic_collective_2018} we used a Maxwell-Liouville approach~\cite{ziolkowski_ultrafast_1995,marskar_multilevel_2011}, beyond the rotating-wave approximation and considering one spatial dimension to investigate the propagation of a FEL pulse through a dense sample of helium ions. For simplicity we do not treat the ionisation step, and assume a purely ionic medium. It is straightfoward to introduce loss terms for diffraction and collisions, but these are neglected here for simplicity. In brief, we use a random phase approximation~\cite{pfeifer_partial-coherence_2010} to generate an approximation to the FEL pulse, using as parameters a pulse length of \SI{70}{\femto\second}, spectral width of \SI{0.5}{\nano\metre}, and pulse energy of \SI{10}{\micro\joule}. Using spatial and temporal stepsizes of \SI{3.0}{\nano\metre} and \SI{6.5}{\atto\second}, the electric field is propagated through an atomic (ionic) region bounded by regions of free space and absorber~\cite{taflove_computational_2005}. All 16 levels of He$^{+}$ with principal quantum number $n\leq4$ were included. While the experimental interaction region is defined by the overlap of the differentially-pumped region (length \SI{5}{\milli\metre}) with the focussed FEL beam (spotsize \SI{10}{\micro\metre}), the ground state neutral atom density is only known to within around an order of magnitude, and the actual distribution of excited ions is likely to vary shot by shot due to differences in absorption, neutral atom density, and pulse energy. Here we consider an idealised case of a fixed length of constant density. The parameters are in a regime where a scaling law~\cite{sun_propagation_2010} can be applied, and the length and density range used (\SI{50}{\micro\metre} $\times$ \SI{2e24}{\per\metre\cubed}) scale to within an order of magnitude of the upper limit estimated experimental conditions (\SI{5}{\milli\metre} $\times$ \SI{2e23}{\per\metre\cubed}). The approach used allows us to study the effects on the incident pulse, and reveals emission at all wavelengths, and in both the forwards and backwards directions.

\begin{figure}
	\includegraphics[width=80mm]{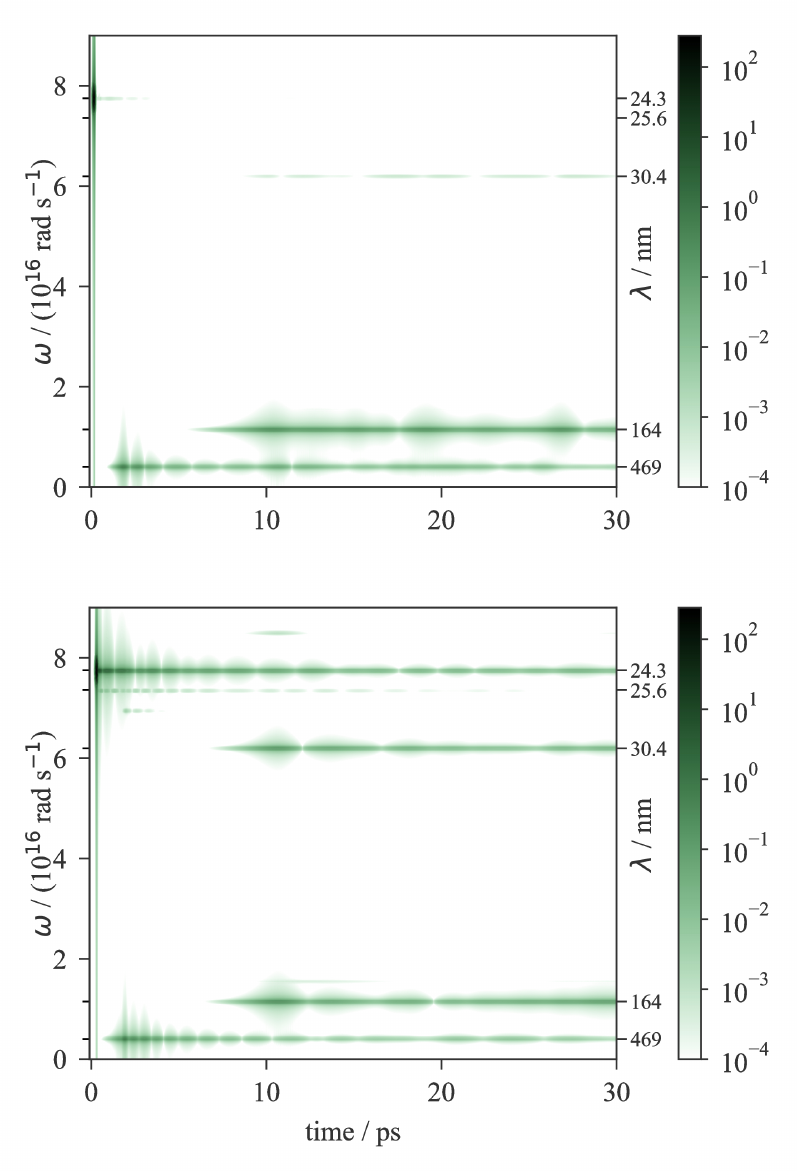}%
	\caption{\label{fig:sim}(Simulation). Rolling Fourier transforms of the electric field at the entrance (upper, backwards emission) and exit (lower, forwards emission) of the ionic medium. The colour scale is consistent between the two plots.
	}
\end{figure}

Figure~\ref{fig:sim} shows results of an example simulation. Plotted (on a log scale) are rolling Fourier transforms (window size \SI{20}{\femto\second}) of the electric field emitted at the exit and entrance of the atomic medium. The spectrally broad FEL pulse can be seen as a thin vertical line centred at $t\approx$ 0 and $\omega$ = \SI{7.8e16}{\radian\per\second}. At the exit of the medium (only), free-induction decay and Burnham-Chiao ringing~\cite{sun_propagation_2010,burnham_coherent_1969} at the resonant wavelength can be seen, with a duration lasting tens of picoseconds. We have directly observed this process experimentally in neutral helium, and these observations will be the subject of a separate publication. After a time delay of around \SI{2}{\pico\second}, pulsed emission is seen at \SI{469}{\nano\metre}, followed several picoseconds later by emission at \SI{164}{\nano\metre}. These emissions occur in both the forwards and backwards directions, and occur earlier for higher number density (simulations at multiple densities confirm a 1/(number density) scaling of pulse delays and widths). In the forwards direction, emission is also seen at \SI{30.4}{\nano\metre}, nearly simultaneous with the emission at \SI{164}{\nano\metre}. This we interpret as \emph{yoked} superfluorescence~\cite{brownell_yoked_1995}, a signature of the coherence with the ground state transferred from the initial excitation. This yoked superfluorescence is expected to depend strongly on the characteristics of the incident pulse, and appears in the simulations only for a small number of realisations of the random phase of the FEL pulses. In contrast the emissions at \SIlist{469;164}{\nano\metre} occur for the majority of realisations. The relative timings of the emissions, and the absence of yoked superfluorescence in the backwards direction are also consistent with previous observations. Prompt emission is also seen in the forwards direction at \SI{25.6}{\nano\metre}, which we interpret as direct excitation and free-induction decay of the 3$p$ state. Four-wave mixing emission is also evident, for example at $\omega$ = $\omega_{\textrm{1}s\textrm{4}p}$ - $\omega_{\textrm{4}p\textrm{3}s}$ + $\omega_{\textrm{3}s\textrm{2}p}$ = \SI{8.50e16}{\radian\per\second} and $\omega$ = $\omega_{\textrm{1}s\textrm{4}p}$ - 2$\omega_{\textrm{4}p\textrm{3}s}$ = \SI{6.95e16}{\radian\per\second}, at the times when the relevant wavelengths are present.
While these simulations do not represent a direct model of the experimental conditions (this would require at least two spatial dimensions to be considered, a more accurate knowledge of number density and spatial distribution, and the inclusion of processes such as collisions), the qualitative agreement strongly backs up our interpretation of the experimental results. 

In summary, we have observed intense, highly directional emission at wavelengths of \SIlist{469;164;30.4;25.6}{\nano\metre} following the ionisation end excitation of helium using FEL pulses at a central wavelength resonant to 4$p$ excitation of He$^+$. We interpret the observations as cascade superfluorescence on the route 4$p$-3$s$-2$p$, yoked superfluorescence on the 2$p$-1$s$ transition, and direct excitation and free-induction decay of the 3$p$ state. The behaviour of the delay times of the upper transition emission is consistent with this interpretation, and a comparison with a full simulation of the propagation of the FEL pulses through a dense medium of helium ions supports our conclusions. Full details of the simulations, and further experimental results from both He$^{+}$ and neutral He will be presented in a forthcoming comprehensive paper.

\begin{acknowledgments}
	The experiments were performed at SACLA BL1 with the approval of the Japan Synchrotron Radiation Research Institute JASRI (proposal numbers 2017A8012, 2017B8083, and 2018A8013). The work was supported by JSPS KAKENHI grants (26286080, 15K04707), and the Research Foundation for Opto-Science and Technology. Calculations were carried out through access to the computing facilities of the Japan Atomic Energy Agency. JH and SK thank Y Miyamoto and N Sasao (Okayama University) for the loan of equipment, and JH the QST SES staff (in particular Y Fukuda) for experimental support. SK was supported by a RIKEN Incentive Research Project. The experiments were made possible by the support of the SACLA engineering team. Mitsuru Nagasono (RIKEN) made important contributions to the early stages of this work, and we thank C Ohae and Y Miyamoto (Okayama University) for helpful input.
\end{acknowledgments}

\end{document}